\documentclass{aa}
\usepackage{graphicx}
\usepackage{txfonts}
\usepackage{natbib}

\bibpunct{(}{)}{;}{a}{}{,}

\begin{document} 

\title{The metal absorption systems of the QSO 0103-260 and the galaxy
redshift distribution in the FORS Deep Field \thanks{Based on
observations obtained with {\sc Uves} at the ESO Very Large Telescope,
Paranal, Chile (proposal No.~66.A--0133).}}

\author{S. Frank\inst{1,2}
           \and I. Appenzeller\inst{2}
           \and S. Noll\inst{2}
           \and O. Stahl\inst{2}}

\offprints{S. Frank}
\institute{Department of Astronomy, Ohio State University, 140 W.18th Ave., 
Columbus, OH 43210, USA
\and 
Landessternwarte, K\"onigstuhl, D 69117 Heidelberg, Germany}

\date{Received 10 February 2003 / Accepted 11 June 2003}

\abstract{Using the {\sc Uves} echelle spectrograph at the ESO VLT, we
observed the absorption line spectrum of the QSO 0103-260 in the {\sc
Fors} Deep Field. In addition to the expected Ly$\alpha$ forest lines,
we detected at least 16 metal absorption systems with highly different
ionization levels in the observed spectral range. The redshifts of the
metal absorption systems are strongly correlated with the redshift
distribution of the high-$z$ galaxies in the {\sc Fors} Deep Field and
with the strength (but not the number density) of the Ly$\alpha$
forest lines. Both the metal systems and the galaxies show clustering
at least up to the QSO emission line redshift of 3.365, but only few
of these galaxy accumulations seem to form bound systems.
\keywords{Galaxies: high redshift - quasars: absorption lines -
intergalactic medium - cosmology: early universe}} \titlerunning{Metal
absorption systems of the FDF QSO 0103-260} \authorrunning{S. Frank et
al.}  \maketitle

\section{Introduction}\label{introduction}

Observations of high-redshift galaxies and of the absorption spectra
of high-redshift QSOs are among the most important sources of
information on the first few Gyrs of the cosmic evolution. Moreover,
since model calculations for the formation of galaxies predict that
galaxies started to form by gravitational contraction of gas
concentrated in the deep potential wells of the dark matter, at high
redshift the intergalactic gas density and the galaxy density are
expected to be closely correlated. The details of this correlation are
expected to yield important information on the formation history of
galaxies and on the interaction of young galaxies with their
environment. As demonstrated recently by \citet{Adelberger:2003}, the
winds and the UV radiation of young galaxies can ionize the
surrounding gas and can thus produce gaps in the Lyman forest line
distribution in the spectra of QSOs. Since atomic nuclei heavier than
those of Lithium cannot have been formed during the Big Bang, the
metal absorption features of the IGM contain valuable information
on the population content and star formation history of the early
galaxies
\citep[see~e.g.][]{2001ApJ...561L.153S,2002A&A...391...21P}. Studies
of IGM absorption lines in the spectra of QSOs and of high-redshift
galaxies along the same line-of-sight promise useful information on
the questions mentioned above. Investigations of this type have been
carried out in the southern Hubble Deep Field (HDF-S) by
\citet{1999A&A...346L..21T} and by \citet{2000AJ....120.1648C}, and in
the direction of 8 high-redshift QSOs by \citet{Adelberger:2003}.  In
the present paper we describe a similar investigation using data from
the {\sc Fors} Deep Field \citep[FDF,
cf.][]{2000Msngr.100...44A,2001RvMA...14..209H,Heidt:2003}.  As
pointed out in the papers quoted above, one of the selection criteria
for the FDF was the presence of a bright high-redshift QSO
\citep[Q~0103-260, cf. ][]{1991ApJS...76...23W,1999A&A...352L...1D}
which was included specifically for the purpose of studying such
correlations. Although in the course of the survey 7 additional QSOs
were detected in the FDF, Q~0103-260 is still the only $z > 2$ quasar
which is bright enough for high-resolution spectroscopy.

The main emphasis of the present investigation is on the correlations 
observed between the galaxy redshift distribution and the redshifts of the
metal line systems of Q~0103-260. However, some information on the
H{\sc i} absorption towards Q~0103-260 is also presented.    

\section{Observations and data reduction}\label{sec_Observations}

The data presented here are based on observations carried out in
service observing mode in November and December 2000 with {\sc Uves},
the {\bf U}ltraviolet and {\bf V}isual {\bf E}chelle {\bf
S}pectrograph at the Nasmyth platform B of ESO's VLT UT2 (Kueyen) on
Cerro Paranal, Chile. For all observations the red arm of {\sc Uves}, 
with a central
wavelength of 5200 \AA\ and a slit width of $1\arcsec $ (resulting in a
usable spectral range of 4166 -- 5164 \AA\ and 5230 -- 6212 \AA\ and
a (measured FWHM) spectral resolution of almost exactly 40\,000),
was used. The
wavelength gap near the central wavelength is due to the gap between
the CCDs of the {\sc Uves} red-arm detector mosaic. Further technical
information on {\sc Uves} can be found in the {\sc Uves} User's Manual
\citep{dodorico:2003} on the ESO Web page.

The total integration time of about 6.9 hours was split into 8
individual exposures of 3100 s each. During most
nights the observing conditions were photometric. 
One exposure was somewhat degraded by
moonlight contamination.  Basic information on the individual frames
is listed in Table~\ref{Observations}. All original spectroscopic
frames are available from the ESO archives in Garching.

\begin{table}
\caption{Observing dates and conditions}\label{Observations}
\begin{center}
\small{
\begin{tabular}{llllll}
\hline
 Date & Start (UT)  & Airm. & Seeing & Remarks\\
\hline
 Nov. 17 & 02:17:38  & 1.00 & 0\farcs5--0\farcs8 & \\
 Nov. 17 & 03:10:22  & 1.03 & 0\farcs5--0\farcs8 & \\
 Nov. 18 & 03:19:15  & 1.05 & 0\farcs4--0\farcs8 & \\
 Nov. 18 & 04:12:13  & 1.15 & 0\farcs4--0\farcs8 & \\
 Nov. 19 & 03:19:53  & 1.06 & 0\farcs6--0\farcs8 & \\
 Nov. 19 & 04:12:44  & 1.17 & 0\farcs6--0\farcs8 & \\
 Dec. 14 & 00:40:34  & 1.00 & 0\farcs6--1\farcs3 & moonlight \\
 Dec. 16 & 02:46:52  & 1.23 & 0\farcs7--0\farcs9 & \\
\hline
\end{tabular}
}
\end{center}
 \end{table}

In order to obtain some information on the spectral region
corresponding to the {\sc Uves} CCD gap, we supplemented our high
resolution {\sc Uves} observations with a medium resolution (measured
FWHM resolution 2\,800) spectrum covering the spectral range 4530 --
5820 \AA\ using {\sc Fors}2 with Grism 1400V at the ESO VLT UT2. The
S/N per resolution element of the medium resolution spectrum is $>
100$.

Since the {\sc Uves} observations were carried out in service mode using
standard settings, the basic spectroscopic reduction procedures (bias
correction, background subtraction, cosmic ray removal, flat-field
correction, order extraction, sky subtraction, wavelength calibration,
merging of the orders and resampling to a common wavelength scale)
were carried out using ESO's {\sc Uves} pipeline reduction procedures.  Flux
standard star observations (with a 10$\arcsec${} slit) were obtained once
during each of the observing nights and were used for a flux
calibration of the reduced data.

The identification and subsequent analysis of the absorption lines was
based on the catalog of \citet{1992ApJS...81..883M}, who lists below
2000 \AA\ vacuum wavelengths. Moreover, the line fitting programme
{\sf VPFIT} was used which assumes vacuum wavelengths as input.  Therefore,
in addition to the heliocentric radial velocity correction, a
transformation from air wavelengths to vacuum wavelengths was applied
to all spectra.  This conversion was carried out following the official
IAU rules (\citet{Oosterhoff:1957}, based on \citet{Edlen:1953}).

\begin{figure} 
\resizebox{\hsize}{!}{\includegraphics[angle=270]{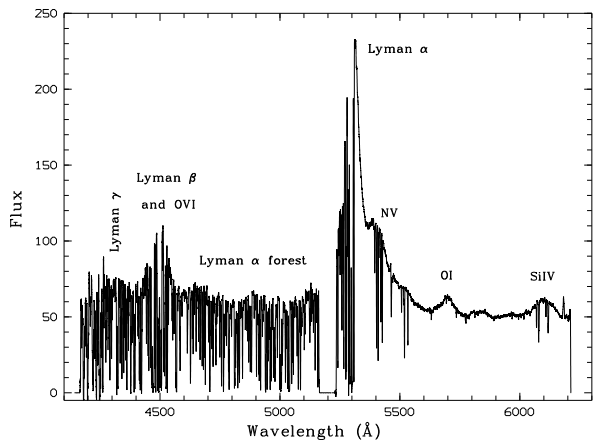}}
\caption[The complete spectrum of Q~0103-260]{The complete {\sc Uves} spectrum 
of Q~0103-260. The gap extending from 5164 \AA\ to 5230 \AA\ resulted from
the space between the two CCDs of the {\sc Uves} red arm CCD mosaic.}
\label{complete_spectrum}
\end{figure}

 
The final spectrum resulting from the co-addition of the individual
reduced spectra and representing a total integration time of 24\,800 s
is displayed in Fig.~\ref{complete_spectrum}. It has a continuum S/N
per resolution element of about 35 -- 50 in the wavelength range 4200
-- 5160 \AA\ and about 50 -- 65 in the range 5230 -- 6210 \AA\@.  The
mean error of the wavelength calibration (as derived from the
dispersion solution) is about 10 m\AA\ in the blue section and about 7
m\AA\ in the red section of the spectrum.

Since this investigation deals with the measurement of absorption
features, some effort was devoted to determine a suitable and
meaningful continuum level and to assess the accuracy of this
function.  As all absorption lines are very narrow compared to QSO
emission features, we were not interested in the true (non-thermal and
thermal) continuum emission of the quasar. Instead, we tried to derive
the quasi-continuum formed by the superposition of the true continua
and the broad emission lines and blends of the QSO\@. As this
quasi-continuum is difficult to predict from models, it was determined
empirically from local 5th-order polynomial fits through apparently
line-free sections of the observed quasi-continuum.  This procedure is
relatively safe in the red spectral region, even at the position of
broad emission lines (except at the peak of the Ly$\alpha$ emission
line, which was omitted from the following analysis).  However, due to
the large number of Ly$\alpha$ absorbers, the assumed quasi-continuum
blue-ward of the Ly$\alpha$ emission line peak depends somewhat on the
subjective judgement of the reality of apparent continuum islands
between the Ly forest lines. In order to estimate the uncertainties
possibly introduced by this subjective judgement, we repeated the
procedure of selecting the fulcrums and the subsequent interpolation
10 times. At positions where the quasi-continuum level appeared
uncertain, we used, on purpose, different assumptions in the different
runs. The 10 versions of the quasi-continuum were then averaged, and
the variance between the different versions was used to estimate the
accuracy of the continuum derivation. We found differences of the
local continuum level derived in this way of up to 7\% in the region
of the Ly$\beta $, O{\sc vi} and Ly$\gamma $ emission lines. In
most parts of the spectrum, however, the deviations were a few percent at
most. Therefore, the uncertainties of the continuum level, while
present, should not have influenced the statistical conclusions
described in this paper. We also note that we checked for, but did not
find any evidence for, a possible under-corrected zero-order
contamination, as described by \citet{1998ARA&A..36..267R}, which
could have influenced our measurements.

\section{Line profile fits}

Although the present study concerns mainly the investigation of the
metal lines in the spectrum of Q~0103-260, it was important to first get
 some information on the Ly$\alpha$ absorption lines which form
the great majority of the absorption features in the observed
spectrum.  According to \citet{1984ApJ...278..486C}, the profiles of
typical Ly$\alpha$ lines in high-resolution (FWHM $<$ 25 km s$^{-1}$)
QSO spectra can normally be approximated by Voigt profiles. Each
line can hence be characterized by its Doppler parameter $b =
\sqrt{2kTm^{-1}_\mathrm{A} + b^{2}_\mathrm{turb}}$, the column density
$N_\mathrm{HI}$, and the redshift $z$.

The line profile $I(\lambda)$ for each line is then given by
\begin{equation}\label{intensity_lambda}
I(\lambda) = I_\mathrm{0}\times{}e^{-\tau{}(\lambda)}
\end{equation}
where the optical depth $\tau$($\lambda$) is 
\begin{equation}\label{optical_depth}
\tau{}(\lambda) =
\frac{N_\mathrm{HI}r_{0}f{}\sqrt{\pi}c\lambda_\mathrm{c}10^{-8}}{b\sqrt{2\ln2}}\times{}V(a,u)
\end{equation}
and $V(a,u)$ is the Voigt profile function. 

A powerful and convenient fit program for Ly forest lines is the code
{\sf VPFIT} which is based on the above assumptions and which has been
developed by \citet{1984ApJ...278..486C}. It is available from
\footnotesize{(www.ast.cam.ac.uk/$\sim$rfc/vpfit.html)}.  \normalsize
This program, which uses $N_\mathrm{HI}$, $b$, and $z$ as fit parameters and
minimizes a suitable $\chi^{2}$ function of these parameters, was used
throughout the present study.

The main application of {\sf VPFIT} was the identification and fit of
the H{\sc i} absorption lines in the region blue-ward of the Ly$\alpha$
emission. As a first approximation, all these lines were assumed to arise from
hydrogen. Hence, the identified metal lines were also included in the
fit. Since {\sf VPFIT} works best with small data
``chunks'', the spectrum was divided into 16 sections of roughly 40 \AA\
overlapping by up to 5 \AA\@. Each section was fitted separately.

Normally, following \citet{2001A&A...373..757K}, more lines should be added
to an existing fit if $\chi^{2}_{\nu}${} $\geq${} 1.3.  In some cases
where regions contained either a saturated line or a multitude of very weak
lines with column densities $N_\mathrm{HI}$ $<$ 10$^{12.5}$ cm$^{-2}$, 
$\chi^{2}_{\nu}$
had to remain greater than the 1.3 threshold.  (Fixing this threshold
obviously is to some extent arbitrary, and increasing the number of
components normally improves the fit. There is, however, no guarantee that a
better fit, i.e.\ a smaller $\chi^{2}_{\nu}$, is always physically
``more correct'').

Tests with different continuum normalizations and with filters applied
to the spectrum before the fitting procedure showed that the fit
results for lines with $\sim 13.2 < \log N_\mathrm{HI} < 15.5$ are very
robust against such differences in the data handling. For very
weak lines ($\log N_\mathrm{HI} < 12.8$) and for saturated lines, the fit
results were found to depend on details of the data reduction. Thus,
the results for such lines should be taken with care and anybody
wishing to make use of our results on very weak or saturated lines
should repeat the data reduction.

\begin{figure}
\resizebox{\hsize}{!}{\includegraphics[angle=270]{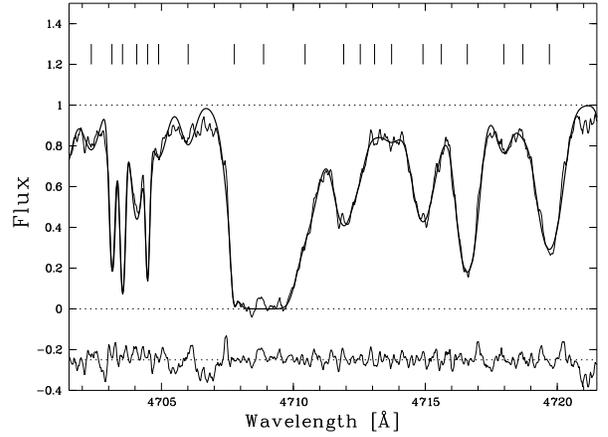}}
\caption[Example of the spectral fits with {\sf VPFIT}]{Example of the
spectral fits obtained with {\sf VPFIT}. The normalised observed
spectrum and the model spectrum (smooth line) are plotted for a 20
\AA\ section  of the Lyman forest.  Also plotted is the difference
(for clarity downshifted by 0.25) between the fit and the data. The
narrow lines at $\sim$ 4703 -- 4705 \AA\ are metal lines ( Si{\sc
iv}$\lambda$1394 at $z \sim$ 2.375 and Fe{\sc ii}$\lambda$2383 at $z
\sim$ 0.974). The tick marks above indicate the positions of the line
centroids of the lines found by the fit procedure. For the saturated
region near 4708 \AA\ (where the fit is ambiguous, see text) assuming
a blend of three lines gave the best result.}\label{example_fit}
\end{figure}

An example of the results of the line fitting procedure with {\sf VPFIT} 
is given in Fig.~\ref{example_fit}.

\section{The Ly$\alpha$ forest lines}

In order to verify that the intergalactic medium in the direction of
the FDF has normal properties, we compared our new data with other
Lyman forest studies of QSOs at similar redshifts.  For this purpose,
we calculated and compared various statistical quantities and
functions, which are briefly described in the following paragraphs.
To avoid confusion with the Ly$\beta$ forest, features caused by
the strong O{\sc vi} lines, and the broad blends close to the
Ly$\alpha$ emission peak, the Ly$\alpha$ forest analysis was
restricted to the wavelength range from 4530 to 5165 \AA\@. In this
wavelength range the fit procedure resulted in the detection of 423
lines. Of these lines, 49 were removed from the sample because their
small line widths ($b <$ 15 km s$^{-1}$) made an identification with
metal absorption more likely, or because a too low ($N_\mathrm{HI}$
$<$ 10$^{12.5}$ cm$^{-2}$) or too high ($N_\mathrm{HI}$ $\geq${}
10$^{16}$ cm$^{-2}$) column density made their analysis uncertain.
Therefore, only 374 ``reliable'' lines were used in the following
analysis of the Lyman forest. However, since the fraction of discarded
lines is still small, the concentration on the ``reliable'' H{\sc i}
lines should not have had any significant influence on the statistical
results.

A plot of the H{\sc i} line fit from 4530 to 5165 \AA\ as well as a line list
derived from this fit is available on the web site
{\footnotesize http://www.astronomy.ohio-state.edu/$\sim$frank}.  The
corresponding catalog lists the redshifts $z_\mathrm{abs}$, the column
densities $N_\mathrm{HI}$, and the broadening parameters $b$ for the
reliably measured Ly$\alpha$ absorption lines.
 
For the 374 lines selected in this way we investigated the following
quantities and relations:
 
\subsection{Column density distribution function}

The column density distribution function (CDDF), $f(N_\mathrm{HI})$,
measures the number of absorption lines per unit column density (in
units of cm$^{-2}$) and per unit absorption path (defined as $x(z) =
1/2\, [(1+z)^2 - 1.0]$ for $q_0 = 0.0$, which was adopted here
primarily for the the reason of comparability to data sets by other
authors) as a function of $N_\mathrm{HI}$. It is comparable to the
luminosity function in the study of galaxies. For the CDDF of the
Ly$\alpha$ forest of Q~0103-260 we find that the distribution can be
well approximated by

$$ f(N_\mathrm{HI}) = 7.1 \times 10^{9} N_\mathrm{H}^{-1.62}$$

which is in good agreement with earlier results demonstrating
that the CDDF can always be approximated by a power law with an
exponent near $-1.5$ \citep{1995AJ....110.1526H,1996ApJ...472..509L,
1997AJ....114....1K,1997ApJ...484..672K,2001A&A...373..757K,
1998ApJ...506....1W,1991ApJ...367...19L}.

\subsection{Line number density}

For the line number density, defined as
$$\frac {dN}{dz} = \left(\frac{dN}{dz}\right)_\mathrm{0}(1+z)^\gamma
$$ where $dN$ is the number of lines per redshift range $dz$,
$(\frac{dN}{dz})_\mathrm{0}$ is the local number density of the forest
($\approx 9.1 \pm 0.4$), and $\gamma \approx 2.19 \pm 0.27$, according
to \citet{2001A&A...373..757K}, we find for $13.6 < N_\mathrm{HI}
< 16.0$:
$$ \log(\frac{dN}{dz}) = 2.27 \pm 0.05 $$ at a mean $z$
of 2.99,  which falls exactly on the corresponding relation derived by
\citet{2001A&A...373..757K} for $z > 1.5$.

\subsection{Mean HI opacity}

For the mean or effective H{\sc  i}
opacity $\tau_\mathrm{eff}$, defined as\\ 
$e^{-\tau_\mathrm{eff}} = <e^{-\tau}>$ 
(where $<>$ indicates averaging over the wavelength) we obtain for 
$z$ = 2.99 the value $0.36 \pm 0.05$, which is again in excellent 
agreement with the results of \citet{2001A&A...373..757K} and others.
  
\subsection{Line counts}

The number of lines in a given $z$ interval as a function of the
flux threshold $F_\mathrm{t}${}, and the number of lines as a function
of the filling factor, defined as the fraction of the spectrum
occupied by pixels whose normalized flux is smaller than a given flux
threshold $F_\mathrm{t}$, again, were found to agree with the
corresponding functions derived in earlier studies
\citep{2001A&A...373..757K,1999MNRAS.310...57S,Outram:1999,
1997ApJ...484..672K, 1995AJ....110.1526H, 1996ApJ...472..509L,
1993ApJ...414...64P}.
 
\subsection{Doppler parameter distribution and temperature-density relation}

The distribution of the Doppler parameter $b$ for the Ly$\alpha$
lines can be well approximated by a Hui-Rutledge distribution
\citep{1999ApJ...517..541H}{} with $b_\mathrm{HR}$ = 13.8 km s$^{-1}$
and $b_{\sigma }$ = ($23.7 \pm 2.8$) km s$^{-1}$.  In order to derive
information on the temperature of the IGM, we investigated the
relationship between the (baryon) density
and the temperature $T$ by looking 
for a cut-off in the observed column density vs. Doppler
parameter distribution. 

The resulting temperatures of $\sim${} 15\,000 -- 20\,000 K at the
mean column density of $N_\mathrm{HI}$ = 1.9 $\times 10^{13}$
cm$^{-2}$  are in good agreement  with photoionization models.

\begin{table}
\caption{Line list for the three metal line systems with redshifts
close to that of the QSO emission lines. N denotes the minimum number
of velocity components observed in the line profile. An asterisk in
the last (Re) column points to a comment in Section 5.2. ``bl'' in
this column indicates that the line is significantly blended.  
}\label{mlinelist1}
\begin{center}
\small{
 \begin{tabular}{clrlrrcl}
\hline
$\lambda _\mathrm{obs}$  & Ident & & $z$ &
N & EW & m.e. & Re\\
(\AA ) & & & & & (\AA )& (\AA ) & \\
\hline
\multicolumn{8}{c}{z3}\\
\hline
 4267.2 & C{\sc iii} &  977 & 3.367 & 4 & $<$.67 & & bl \\
 4506.8 & O{\sc vi}  & 1032 & 3.367 & 4 & $<$2.0 &  & bl\\       
 4531.7 & O{\sc vi}  & 1038 & 3.367 & 4 & $<$2.0 &  & bl\\ 
 5269.3 & Si{\sc iii} & 1207 & 3.367 & 4 & $<$.68 & .038 & *\\
 5410.4 & N{\sc v}   & 1239 & 3.367 & 4 & 1.503 & .032 & \\           
 5427.8 & N{\sc v}   & 1243 & 3.367 & 4 & 1.222 & .020 & \\
 5827.9 & C{\sc ii}  & 1335 & 3.367 & 4 & 0.032 & .015 & \\ 
 6086.6 & Si{\sc iv} & 1394 & 3.367 &     & $<$0.05 & & \\
 6125.9 & Si{\sc iv} & 1403 & 3.367 &     & $<$0.05 & & \\
\hline
\multicolumn{8}{c}{z2}\\
\hline
 4262.8 & C{\sc iii} &  977 & 3.363 & 4 & $<$2.6 & & bl \\ 
 4318.5 &  N{\sc iii} &   990 & 3.363 & 7 & 0.329 & .060 & \\
 4502.4 &  O{\sc vi} &  1032 & 3.363 & 7 & $<$2.9 &  & bl \\      
 4527.2 &  O{\sc vi} &  1038 & 3.363 & 7 & $<$3.8 &  & bl \\      
 5264.1 &  Si{\sc iii} & 1207 & 3.363 & 7 & 1.190 &.066 & \\
 5405.1 &  N{\sc v}  &  1239 & 3.363 & 7 & 2.072 & .057 &  \\              
 5422.5 &  N{\sc v}  &  1243 & 3.363 & 7 & 1.444 & .061 & \\
 5422.6 & C{\sc ii}  &  1335 & 3.363 &   & $<$.04 &  & \\             
 6081.1 &  Si{\sc iv}&  1394 & 3.363 & 7 & 1.280 & .072 & \\              
 6120.3 &  Si{\sc iv}&  1403 & 3.363 & 7 & 0.906 & .065 & \\              
\hline
\multicolumn{8}{c}{z1}\\
\hline
 4256.0 &  C{\sc iii} &  977 & 3.356 & 4 & 0.920 & .029 &  \\
 4494.9 &  O{\sc vi}  & 1032 & 3.356 & 5 & $<$4.0 &  & bl \\      
 4519.8 &  O{\sc vi}  & 1038 & 3.356 & 5 & $<$4.2 &  & bl \\
 5255.6 &  Si{\sc iii} & 1207 & 3.356 & 5 & 0.460 & .035& \\
 5396.4 &  N{\sc v}   & 1239 & 3.356 & 5 & 0.490 & .042 & \\              
 5413.8 &  N{\sc v}   & 1243 & 3.356 & 5 & 0.290 & .037 & \\
 5813.2 & C{\sc ii}   & 1335 & 3.356 &  & $<$.03 &   & \\               
 6071.2 &  Si{\sc iv} & 1394 & 3.356 & 5 & 0.430 & .028 & \\              
 6110.5 &  Si{\sc iv} & 1403 & 3.356 & 5 & 0.290 & .030 & \\              
\hline
\end{tabular}
}
\end{center}
\end{table}

\begin{center}
\begin{table}
\caption{Line list for the metal absorption line systems
with $2.1 < z < 3.2$}\label{mlinelist2}
\begin{tabular}{clrlrrcl}
\hline
$\lambda _\mathrm{obs}$  & Ident & & $z$ &
N & EW & m.e. & Re\\
(\AA ) & & & & & (\AA )& (\AA ) & \\     
\hline                  
 4300.5 &  O{\sc vi}  & 1038 & 3.145 & 2 & 0.045 & .020 & *\\
 5134.6 &  N{\sc v}   & 1239 & 3.145 & 2 & 0.047 &.021 & *\\      
 5151.1 &  N{\sc v}   & 1243 & 3.145 & 2 & 0.041 & .020 & *\\
 5531.2 &  C{\sc ii}  & 1335 & 3.145 & 2 & 0.019 & .020 & *\\ 
 5776.7 &  Si{\sc iv} & 1394 & 3.145 & 2 & 0.194 &.029 &\\  
 5814.2 &  Si{\sc iv} & 1403 & 3.145 & 2 & 0.130 & .025 &  \\              
        &      &      &       &     &          &    & \\
 4271.2 &  O{\sc vi} &  1038 & 3.116 & 2 & 0.029  & .020 & *\\
 4966.5 &  Si{\sc iii} &1207 & 3.116 & 2 & 0.347 &  .035 & *\\ 
 5115.6 &  N{\sc v}  &  1242 & 3.116 & 2 & 0.021 & .012 & *\\         
 5737.3 &  Si{\sc iv} & 1394 & 3.116 & 2 & 0.154 & .023 & \\            
 5774.4 &  Si{\sc iv} & 1403 & 3.116 & 2 & 0.120 & .022 & \\            
        &     &      &       &     &          &    & \\
 4557.4 &  N{\sc v}   & 1239 & 2.679 & 1 & $<$0.04 & 0.015 & *\\             
 5695.6 &  C{\sc iv}  & 1548 & 2.679 & 1 & 0.105 & .013 & * \\            
 5705.0 &  C{\sc iv}  & 1551 & 2.679 & 1 & 0.063 & .013& *\\            
        &      &      &       &     &          &    & \\
 5508.6 &  C{\sc iv}  & 1548 & 2.558 & 1 & 0.268 & .023 &\\             
 5517.8 &  C{\sc iv}  & 1551 & 2.558 & 1 & 0.169 & .022 &\\             
       &      &      &       &     &          &    & \\
 4958.3 &  Si{\sc iv} & 1394 & 2.558 & 1 & 0.042 & .020 & \\            
 5507.8 &  C{\sc iv}  & 1548 & 2.558 & 1 & 0.155 & .021 & \\            
 5517.0 &  C{\sc iv}  & 1551 & 2.558 & 1 & 0.090 & .013 & \\            
       &      &      &       &     &          &    & \\                       
 4703.3 &  Si{\sc iv} & 1394 & 2.375 & 2 & 0.430 & .030 & \\ 
 4733.7 &  Si{\sc iv} & 1403 & 2.375 & 2 & 0.385 & .040 &  \\
 5224.4 &  C{\sc iv}  & 1548 & 2.375 & 2 & 0.350 & .068 & * \\   
        &       &      &       &   &       &      &   \\
      &      &      &       &     &          &    & \\
 4330.6 &  C{\sc ii}  & 1334 & 2.245 & 2 & 0.188 & .010 & *\\            
 4522.8 &  Si{\sc iv} & 1394 & 2.245 & 2 & 0.350 & .020 & *\\ 
 4552.0 &  Si{\sc iv} & 1403 & 2.245 & 2 & 0.290 & .015 & *\\
 4954.2 &  Si{\sc ii} & 1527 & 2.245 & 2 & 0.068 & .025 & *\\             
 5023.9 &  C{\sc iv}  & 1548 & 2.245 & 2 & ? & & bl\\      
 5032.3 &  C{\sc iv}  & 1551 & 2.245 & 2 & ? & & bl\\
 6018.6 &  Al{\sc iii} & 1855 & 2.245 & 2 & 0.109 & .010 & \\
 6044.8 &  Al{\sc iii} & 1863 & 2.245 & 2 & 0.045 & .010 & \\
      &      &      &       &     &          &    & \\
 4329.6 &  C{\sc ii}  & 1334 & 2.244 & 2 & 0.042 & .014 & \\            
 4521.7 &  Si{\sc iv} & 1394 & 2.244 & 2 & 0.112 & .020 & \\ 
 4551.0 &  Si{\sc iv} & 1403 & 2.244 & 2 & $<$.11 &  & bl\\             
 5022.7 &  C{\sc iv}  & 1548 & 2.244 & 2 & $<$.19 & & bl\\      
      &      &      &       &     &          &    & \\
 4421.8 &  Si{\sc iv} & 1394 & 2.173 & 1 & 0.133 & .010 & \\             
 4450.5 &  Si{\sc iv} & 1403 & 2.173 & 1 & 0.095 & .011 & \\
 4911.8 &  C{\sc iv}  & 1548 & 2.173 & 1 & 0.360 & .020 & \\
 5884.2 &  Al{\sc iii}  & 1855 & 2.173 & 1 & 0.032 & .015 & \\
 5909.6 &  Al{\sc iii}  & 1863 & 2.173 & 1 & 0.042 & .016 & \\          
       &      &      &       &     &          &    & \\
 4910.8 &  C{\sc iv}  & 1548 & 2.172 & 1 & 0.180 & .015 & *\\
 4919.0 &  C{\sc iv}  & 1551 & 2.172 & 1 & 0.157 & .016 & *\\          
       &      &      &       &     &          &    & \\
 4836.2 &  C{\sc iv}  & 1548 & 2.124 & 1 & 0.168 & .010 & \\           
 4844.2 &  C{\sc iv}  & 1551 & 2.124 & 1 & 0.122 & .012 & \\
 5793.6 &  Al{\sc iii}  & 1855 & 2.124 & 1 & 0.031 & .019 & \\
 5818.8 &  Al{\sc iii}  & 1863 & 2.124 & 1 & 0.015 & .015 & \\      
\hline
\end{tabular}
\end{table}
\end{center}

\begin{table}
\caption{Line list for the two metal absorption line systems
near $z$ = 0.974}\label{mlinelist3}
\begin{center}
\small{
 \begin{tabular}{clrlrrcl}
\hline
$\lambda _\mathrm{obs}$  & Ident & & $z$ &
N & EW & m.e. & Re\\
(\AA ) & & & & & (\AA )& (\AA ) & \\     
\hline
 4688.1 &  Fe{\sc ii} & 2374 & .9744 & 5 & 0.260 & .080 & bl\\
 4704.5 &  Fe{\sc ii} & 2383 & .9744 & 5 & 0.399 & .030 & bl\\
 5107.0 &  Fe{\sc ii} & 2587 & .9744 & 5 & 0.240 & .035 & bl\\   
 5133.7 &  Fe{\sc ii} & 2600 & .9744 & 5 & 0.438 &.030 & bl\\ 
 5521.1 &  Mg{\sc ii} & 2796 & .9744 & 5 & 1.080 &.040 & \\             
 5535.2 &  Mg{\sc ii} & 2804 & .9744 & 5 & 0.880 & .030 & \\ 
 5632.8 &  Mg{\sc i}  & 2853 & .9744 & 5 & 0.235 & .030 &\\  
      &      &      &       &     &          &    & \\
 4626.8 &  Fe{\sc ii} & 2344 & .9737 & 3 & 0.020 & .030 & bl\\  
 4702.9 &  Fe{\sc ii} & 2383 & .9737 & 3 & $<$0.130 & &*\\
 5132.0 &  Fe{\sc ii} & 2600 & .9737 & 3 & 0.026 & .050 &\\
 5519.2 &  Mg{\sc ii} & 2796 & .9737 & 3 & 0.229 & .030 & \\  
 5533.4 &  Mg{\sc ii} & 2804 & .9737 & 3 & 0.120 & .030 & \\
 5631.9 &  Mg{\sc i}  & 2853 & .9737 & 3 & 0.030 & .030 & \\  
\hline
\end{tabular}
}
\end{center}
\end{table}

From the above comparisons we conclude that the
Ly$\alpha$ forest of the FDF quasar has parameters which are rather typical 
for this redshift and that the IGM along the line of sight to Q~0103-260  
has normal properties.

\section{The metal lines}

\subsection{Line identifications}

About 10~\% of the absorption lines in the {\sc Uves} spectrum belong
to UV transitions of common ions of elements heavier than He in
various ionization states.  Following astrophysical traditions, all
such lines are called ``metal lines'' in this paper. Except for very
weak and very strong lines, the metal lines in the {\sc Uves} spectrum
are easily recognized on the basis of their lower line width (see
e.g. Fig.~\ref{example_fit}). As suggested by
\citet{1997ApJ...489....7R} we assumed all reasonably strong lines
with $b <$ 15 km s$^{-1}$ to be metal lines.  In order to assign these
lines to certain atoms, ions, and redshifts, we used the following
procedure: First, we identified the absorption lines connected with
the QSO and its immediate environment. From earlier studies of the
{\sc Fors} Deep Field it is known that Q~0103-260 is embedded in an
accumulation of bright starburst galaxies at nearly the same redshift
\citep[][]{Heidt:2003a}. Hence, we expected (and found) strong
absorption line systems superimposed on the quasar emission
lines. Because of their association with the quasar emission lines and
their known redshifts (within 0.01 of the emission line redshift)
these lines could be readily identified independently of their
width. In total we found three different groups of lines with $z
\approx z_\mathrm{QSO}$. In Table~\ref{mlinelist1} these redshift
groups are labeled (in the order of increasing redshift) as z1, z2,
and z3.
   
As a next step, we used our identified Ly$\alpha$ forest lines to look 
for metal emission at the redshifts of the strongest 
unblended H{\sc i} absorption
lines. The rest wavelengths and expected line strengths of the metal lines
were taken from the catalog of \citet{1992ApJS...81..883M}. For those 
narrow lines which were still unidentified after this step, we 
calculated mutual wavelength ratios which were compared with those expected
for common Ly forest metal lines (making use of the fact
that wavelength ratios are invariant to the redshift). At the high
spectral resolution of {\sc Uves}, this method is 
particularly powerful for identifying lines belonging to multiplets. 
For each line we then
calculated the redshift. A metal absorption system was assumed to
be established when at least 3 lines or blends (in some cases 
including an H{\sc i} line) were found to have redshifts not deviating by
more than 0.0001. Having established a metal system in this way, we 
searched for other expected lines at the same redshift. 

Using this procedure, we were able to identify more than 80 metal
lines or blends belonging to 16 redshift systems. About 20 (mostly
weak) sharp lines remained unidentified. Lines were assigned to the
same redshift system if their profiles overlapped for at least one
metal line.  With few exceptions each system contains several
components.  In some cases an optimal line fit required the assumption
of up to 16 components. However, as pointed out above, optimized line
fits may overestimate the number of real components. Therefore, we
determined for each system also a lower limit of components by
counting the profile components which were directly visible on our
spectra. The metal lines and blends, for which reliable observed
wavelengths or equivalent widths could be measured, are listed in
Tables~\ref{mlinelist1} to~\ref{mlinelist3}. Lines and blends
belonging to the same system are grouped together. In cases where
only one component of a doublet is listed, the second component could
not be measured because of blending. But in all these cases it has
been verified that the spectrum is consistent with the presence of
both components at the expected absorption strength ratio. The
individual columns of the tables give the mean observed wavelength of
the line (averaged over the individual velocity components), the line
identification, the approximate mean redshift $z$, the lower limit of
velocity components $N$, the total observed equivalent widths EW
(derived by summing up the contributions of all individual velocity
components), and their mean errors.  An asterisk in the last column
indicates a comment in the Subsection 5.2., whereas ``bl'' denotes
blending with other lines (usually H{\sc i}). For some lines blending
made it impossible to derive reliable equivalent widths.  We then give
as upper limits the EW which the line would have if the contribution
of other lines were negligible.  In addition to data for reliably
detected lines, some upper limits or non-significant EW values have
been included in the tables if these values provide information on
line ratios important for the derivation of the ionization parameter
or the plasma density. For many redshift systems, additional lines are
visible in the {\sc Uves} spectrum, but cannot be measured because of
blending. This applies in particular to wavelengths below 4500 \AA\
where the Ly$\alpha$ and Ly$\beta $ forests merge.
     
\subsection{Remarks on individual metal absorption systems}

\subsubsection*{$z$ = 3.367 (z3):}

At least part of the absorption at the expected position of the Si{\sc
iii} resonance line seems to be due to Ly$\alpha$. Therefore, only an
upper limit for the Si{\sc iii} absorption can be given.  Since Si{\sc
iv} is absent in this system, and since the strength of O{\sc vi} and
N{\sc v} suggest a very high ionization parameter, a significant
Si{\sc iii} absorption is not to be expected, but cannot be ruled out
by our observations.

\subsubsection*{$z$ = 3.145:}

All metal lines of this system are weak. Only the lines of Si{\sc iv} are
significant detections. 

\subsubsection*{$z$ = 3.116:}

Weak O{\sc iv} and N{\sc v} and strong Si{\sc iii} indicate a low
ionization state. C{\sc ii}, however, is not detected. Si{\sc iii}
could be overestimated due to blending with a weak
H{\sc i} line.

\subsubsection*{$z$ = 2.679:} 

Except for the three lines listed, all
lines coincide with regions of strong and complex blends.  

\subsubsection*{$z$ = 2.375:} 

EW(C{\sc iv}$\lambda$1548) has been measured in the {\sc Fors}
spectrum. N{\sc v}$\lambda$1243 is clearly detected, but not measurable because
of too severe blending.

\subsubsection*{$z$ = 2.245:} 

The system consists of more than two (at least 3, possibly 7) components. 
The wavelength and EW values refer to the two strongest and best defined
components only. The line ratios indicate a much lower ionization
parameter than observed in the other systems.

\subsubsection*{$z$ = 2.172:}

The only statistically significant metal lines are those of the C{\sc iv} resonance 
doublet. Si{\sc iv} is not detected, but the well defined sharp lines
and the excellent agreement with the theoretical wavelength ratio
makes it very likely that the identification of this weak system
is correct.  

\subsubsection*{$z$ = 0.9737:} 

The Fe{\sc ii}$\lambda$2383 \AA\ line could not be measured accurately
because of an uncertain continuum level at the corresponding
wavelength. The strong lines of low ionization stages indicate that
the ionization parameter is much lower than in the systems with higher
redshifts.

\subsection{General properties of the observed metal systems}

\subsubsection{The metal absorbers with z$_\mathrm{abs}\sim$z$_\mathrm{em}$(QSO)}

As mentioned above and listed in Table~\ref{mlinelist1}, there are
three strong absorption systems with lines of C{\sc iii}, N{\sc iii},
N{\sc v}, O{\sc vi}, Si{\sc iii}, and Si{\sc iv}, at redshifts close
to the quasar's emission line redshift \citep[$z$ = 3.365, according
to][]{Noll:2002}. Such systems with $z_\mathrm{abs} \approx
z_\mathrm{em}$, which are observed in many quasar spectra, and
are often referred to as the ``intrinsic'' absorbers, have been
proposed in the literature to be either due to matter ejected from the
QSOs or to originate in galaxies physically associated with the
quasars.  In a few cases evidence for a close association of such
systems with the central QSO has been reported, e.g.\ time-variability
of the line strength \citep{1995ApJ...443..606H} or partial coverage
of the background source
\citep{1994A&A...291...29P,1999AJ....117.2594G}. In our case,
the known accumulation of galaxies with about the same redshift
around Q~0103-260 \citep[e.g.][]{2001RvMA...14..209H} suggests
that galaxies associated with the QSO at least contribute.  On the
other hand, of the three galaxies with reliable spectroscopic
redshifts $z \approx z_\mathrm{QSO}$ and closer than 5 arcsec to the
line of sight to the QSO only one \citep[object 9003 in][]{Noll:2002}
has a redshift coinciding within the error limits with one of the
systems listed in Table~\ref{mlinelist2} (z3), while the two other
galaxies have slightly larger (by 0.008--0.013) redshifts, indicating
that they are located behind the QSO\@.  However, there are additional
faint galaxies close to the QSO without known spectroscopic redshifts,
which could well cause the other $z \approx z_\mathrm{QSO}$ absorption
systems.

For the system z3 we find a slightly (by about 0.002) higher redshift
than derived for the QSO emission lines. However, as this difference
is within the errors of the derivation of the emission line redshift,
this result cannot be regarded as evidence for a motion of the
absorbing gas towards the QSO\@.

\begin{center}
\begin{figure}
\resizebox{\hsize}{!}{\includegraphics{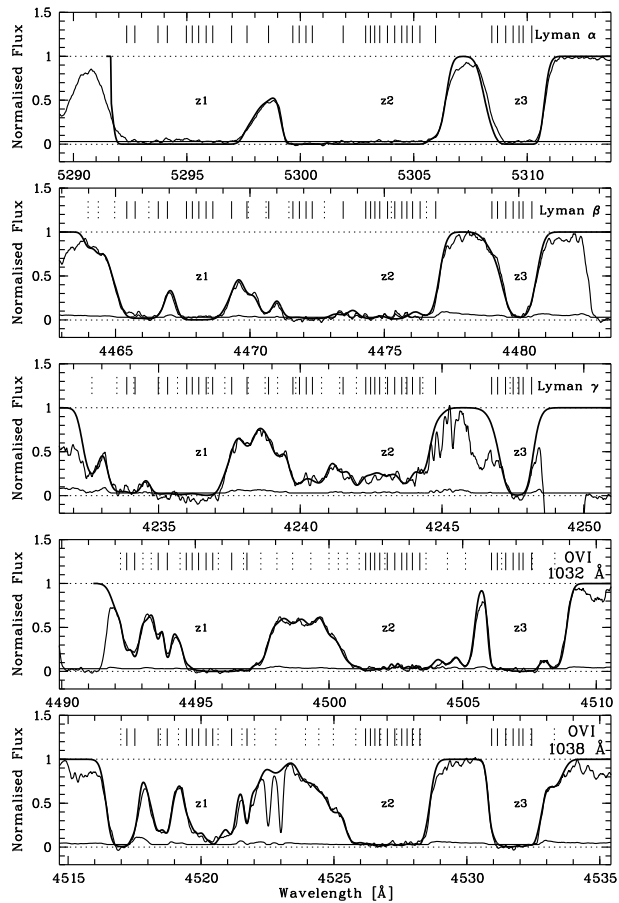}}
\caption[The model for the three ``intrinsic'' absorbers I]{Line
profiles of the blends produced by the metal absorption systems z1 to
z3 with redshifts close to the quasar emission line redshift (thin
line). Also included is the the model spectrum produced by the fit
procedure (smooth thick line) and the mean error of the observed
spectrum (thin line close to the zero level).  The solid tick marks
above the spectra indicate the centroids of the different line
components assumed for the fit. Dashed tick marks represent the
centroids of H{\sc i} at other redshifts added to obtain adequate
fits.  The relatively large residuals near Ly$\gamma$ are due to a
detector defect (at 4249\AA) and the omission of lower-$z$ Ly$\alpha$
lines in the fit. }\label{plot_z123_1}
\end{figure}
\end{center}

\begin{center}
\begin{figure}
\resizebox{\hsize}{!}{\includegraphics{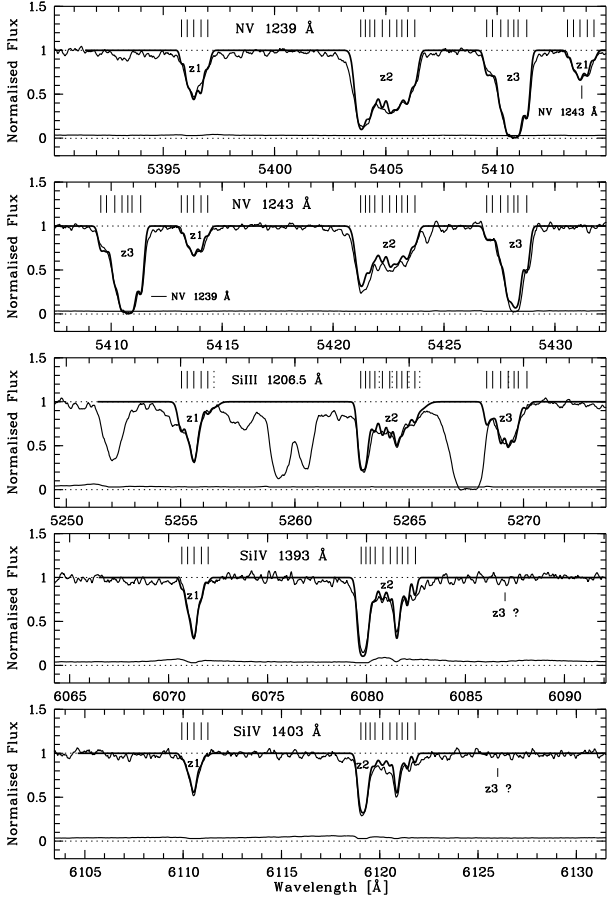}}
\caption[The model for the three ``intrinsic'' absorbers II]{Line
profiles of the blends produced by the metal absorption systems z1 to
z3 with redshifts close to the quasar emission line redshift,
continued. The H{\sc i} blends between the three Si{\sc iii}
features could not be fitted unambiguously and, therefore, were
excluded from the fit.}\label{plot_z123_2}
\end{figure}
\end{center}

In Figs.~\ref{plot_z123_1} and~\ref{plot_z123_2} the profiles of the
three intrinsic absorption systems z1 to z3 and the corresponding
spectral fits are presented. For these fits it was assumed that
the same velocity components with exactly the same individual
redshifts were present at all lines, while the absolute and relative
strength of the components was allowed to vary from line to line.  The
wavelengths of the individual components are, if present, indicated in
Figs.~\ref{plot_z123_1} and~\ref{plot_z123_2}. A table of the fit
parameters and the column densities derived from the fits is also
available electronically from the web page noted above. It is evident
from the plotted profiles that each system consists of discrete
absorptions with different velocities and that the different velocity
components have different strengths for different lines.  As
demonstrated by the figures and Table~\ref{mlinelist1}, all three
systems show very strong high-ionization lines while significant
absorption features of low-ionization stages (like the resonance lines
of C{\sc ii}, N{\sc ii}, Si{\sc ii}, and O{\sc i}) were not detected.
Si{\sc iv} is fairly strong in z1 and z2, but not detected at z3.  On
the other hand, an absorption feature is present at the expected
position of the Si{\sc iii} line of the z3 system. However, the
profile of this feature indicates that at least part (and possibly
all) of this feature is due to Ly$\alpha$. A comparison with
photoionization models
\citep[see~e.g.][]{1986A&A...169....1B,1990ApJS...74...37S,
1996AJ....112..335S} indicates that the ionization parameter $\Gamma $
(while high for all three ``intrinsic'' systems) increases from z1 to
z3. Hence, it appears likely that the apparent Si{\sc iii} line of z3
is in fact due to H{\sc i} at a lower redshift and that Si{\sc iv}
(and lower Si ionization stages) have been destroyed by strong
radiation short-ward of the He$^{+}$ break.

Because of the relatively few significant line strength measurements
our data do not allow to derive reliable quantitative values for the
densities, ionization parameters and abundances in the observed metal
systems. From the EW ratio N{\sc v}/Si{\sc iv} it is clear, however,
that the lines are formed in a plasma of relatively low density and a
very high ionization parameter, and that the ionization parameter is
decreasing with decreasing redshift. Such a behavior is to be expected
if the redshift is related to the distance and if the QSO provides the
ionizing radiation.

\subsubsection{The metal absorbers with 2.0 $< z <$ 3.2}

Compared to the ``intrinsic'' systems all observed metal systems at lower
redshifts have line ratios indicating lower ionization parameters.
Most of these systems, though, are dominated by high-ionization lines too,
with lines from neutral or single ionization stages not being detectable.
Only one system ($z$ = 2.245) shows significant absorption by
low-ionization lines. 

The high state of ionization of the metal systems listed in
Table~\ref{mlinelist2} agrees well with the theoretical predictions of
\citet{1997ApJ...481..601R} and \citet{1997euvl.conf.....B} who used
hydrodynamic computations to calculate the spatial distribution and
line absorption of low-metallicity intergalactic gas at high
redshift. From a detailed comparison of the relative line strengths in
Table~\ref{mlinelist2} with the calculated column densities of
\citet{1997ApJ...481..601R}, we find the high-ionization lines (such as
O{\sc vi}, N{\sc v}, C{\sc iv}) to follow the predictions rather
well. The theoretical results of \citet{1997ApJ...481..601R} also
provide a plausible explanation for the weakness of absorption lines
of low ionization species such as C{\sc ii}, which are expected to be
present only in the rare cases where the line-of-sight passes through a
high density core of a protogalactic clump of the intergalactic gas.

The agreement with the predictions of \citet{1997ApJ...481..601R} is
less good for species with intermediate ionization potentials. In
particular, we find the Si{\sc iv} and Si{\sc iii} lines listed in
Table~\ref{mlinelist2} to be on average stronger and the resulting
Si{\sc iv}/C{\sc iv} and Si{\sc iv}/N{\sc v} column density ratios
about 10 times higher than predicted for a metal-poor gas with a solar
relative composition. As in the case of the measurements of
\citet{1996AJ....112..335S}, the discrepancy is smaller if our data are
compared to the calculations made assuming the chemical composition of
galactic metal-poor stars. But even in this case, our Si{\sc iv} lines
are on average too strong by about a factor of 4.

A comparison with \citet{2002AJ....123.1847M} indicates that the
observed number density of C{\sc iv} systems with EW $>$ 0.15 \AA\ in
the redshift range $2.1 < z < 3.2$ along the line-of-sight to
Q~0103-260 is normal for a QSO at this redshift. Although additional
systems with smaller EW values are present in Table~\ref{mlinelist2},
our data provide no evidence for a high abundance of weak metal
systems.

\subsubsection{The Mg/Fe{\sc ii} systems at $z \approx 0.974$}

\begin{figure}
\resizebox{\hsize}{!}{\includegraphics{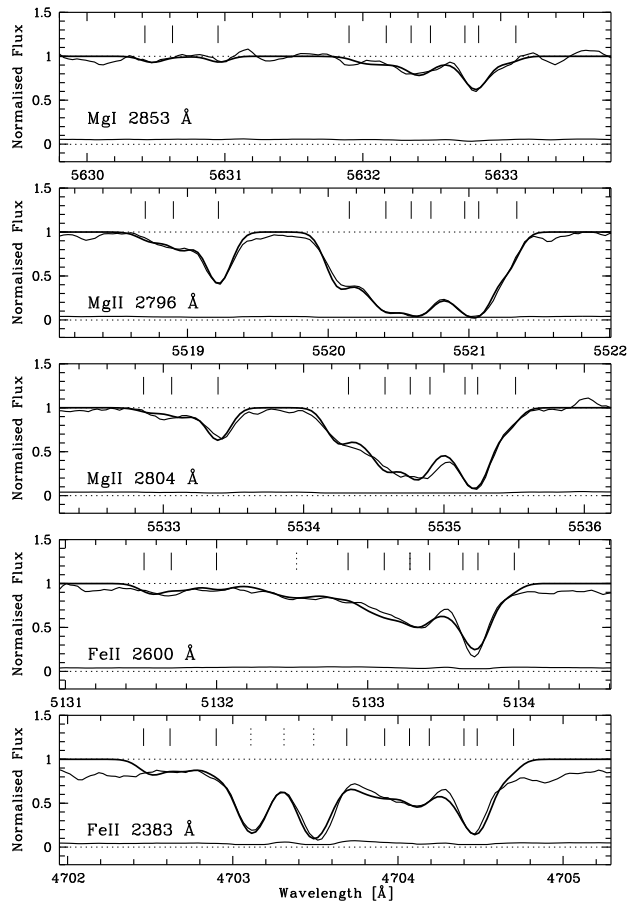}}
\caption[The model for the Mg{\sc i}/{\sc ii} - Fe{\sc ii} absorber at
$z$ = 0.974]{The two Mg/Fe{\sc ii} absorption systems at $z \approx
0.974$. Also plotted is the model spectrum (thick smooth line). The
tick marks again indicate the positions of the centroids of the lines
assumed for the fit. The two strong absorption lines at 4703.1 \AA\
and 4703.5 \AA\ in bottom panel are due to Si{\sc iv}$\lambda$1394 at
$z$ = 2.375.}\label{FeII_1}
\end{figure}

While there are several sharp lines in the red part of the spectrum
which are suspected to be due to low-ionization lines at moderate
redshifts, only two metal systems with redshifts $<$ 2 could be
reliably identified.  The profiles of the strongest lines of these two
systems with almost identical redshifts ($z \approx $ 0.974) are
presented in Fig.~\ref{FeII_1}.  Both systems consist of various
components and show a complex structure.  The plasma producing these
lines has a relatively low ionization level, comparable to that of
most of the interstellar matter in our galaxy and the gas observed in
low-$z$ metal absorption systems \citep[see
e.g.][]{1991A&A...243..344B}.

\section{Correlations}

\begin{figure}
\resizebox{\hsize}{!}{\includegraphics[angle=270]{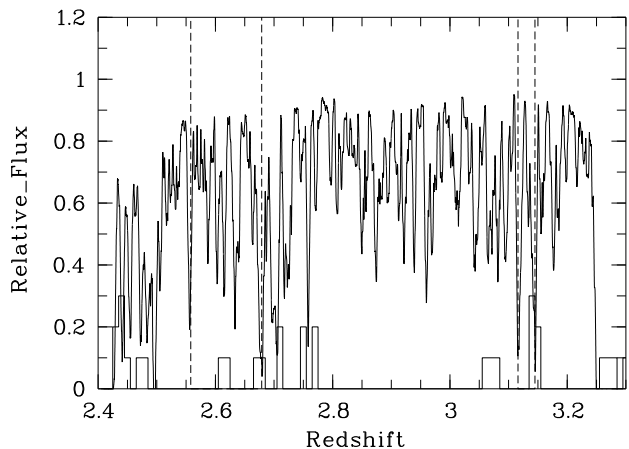}}
\caption[Positions of FDF galaxies with 2.5 $<$ z $<$ 3.5]{The
normalized flux in the QSO spectrum showing the strength of the
Ly$\alpha$ absorption as a function of redshift.  Also given are the
redshift positions of the metal absorption systems in this redshift
range (vertical broken lines) and the histogram of the redshift
distribution of the FDF galaxies (scaled by a factor of 0.1). For this
figure the {\sc Uves} spectrum has been smoothed to a resolution of $\Delta z
= 0.005$.  It has been normalized to the continuum level of the
unsmoothed spectrum.  Note that the strong broad absorption features
at $z$ = 2.70 and 2.72 are in fact due to O{\sc vi} absorption at $z$ =
3.36. The Ly$\alpha$ absorption at 2.68 is contaminated by Ly$\beta $
at $z$ = 3.36 and Ly$\gamma $ of this redshift contributes to the
feature at 2.49. Most other strong features are due to Ly$\alpha$.
}\label{FDF_zdistribution_2.5_3.5}
\end{figure}

Because of the relatively small observed Ly$\alpha$ redshift range and
the presence of the superimposed metal lines, the distribution of the
Ly$\alpha$ absorption-line redshifts provides only limited information
on the cosmic structure along the line-of-sight to the observed
quasar. In particular, we found for the redshift range 2.7 $<$ z $<$
3.3, where we have the most reliable data on the Ly$\alpha$
absorption, a number density distribution of the Ly$\alpha$ lines
which is almost fully consistent with a random distribution of the
absorbers along the LOS\@.  A clustering signal could be detected only
for the two-point correlation function in the lowest velocity range
($\Delta v \leq${} 50 km s$^{-1}$) and in the step optical depth
function \citep{1998ApJ...496..577C} for velocity separations $\sim${}
5--10 km s$^{-1}$.  The clustering signal is higher for the systems
with higher column density. This result is in good agreement with
earlier reports on the clustering of Lyman forest lines on small
scales \citep[see e.g.][]{Webb:1987,1995AJ....110.1526H,
1996ApJ...472..509L,
1995ApJ...440..431C,1995MNRAS.273.1016C,1997MNRAS.285..209C}. However,
while little structure is evident in the number density distribution
of the Ly$\alpha$ absorbers, structure is indicated in the total H{\sc
i} absorption as a function of redshift
(cf. Fig.~\ref{FDF_zdistribution_2.5_3.5}). More specifically, we find an
absence of high column density absorption and a lower-than-average
total Ly$\alpha$ absorption in the redshift interval $2.8 < z < 3.1$.

More interesting information on the distribution and the properties of
the absorbing gas along the LOS to Q~0103-260 is provided by
the redshift distribution of the metal-line systems and by a
comparison of this distribution with the galaxy redshift distribution
in the FDF\@. From Tables~\ref{mlinelist1} to~\ref{mlinelist3} it is
obvious that the metal absorption systems also tend to cluster on
small scales.  Moreover, no metal absorption system is detected in the
redshift range $2.8 < z < 3.1$ where the Ly$\alpha$ absorption is
weak. A comparison between the redshifts of the individual metal absorption
systems and those of the Ly$\alpha$ absorption lines shows that (as
expected) detected metal line systems always correspond to strong
Ly$\alpha$ absorption. On the other hand, there are strong Ly$\alpha
$ absorption features without detectable corresponding metal
absorption. A notable example is the Ly$\alpha$ absorption at $z$ =
2.758. The total Ly$\alpha$ absorption of this system is similar to or
slightly stronger than that for the two systems at $z$ = 2.558, but no
metal absorption is detectable. The upper limits for the C{\sc iv} and
Si{\sc iv} absorption are by factors of at least 0.05 to 0.1 lower
than the corresponding metal line strengths in the $z$ = 2.558
systems. We regard these large differences as clear evidence
for strong local variations of the metal enrichment at this redshift.

The existence of photometric and spectroscopic redshifts for a
significant fraction of the FDF galaxies allows us to compare the
redshift distributions of the galaxies with those of the absorption
line systems discussed here.  At present photometric redshifts are
known for about 6500 FDF galaxies
\citep{2001defi.conf...96B,Gabasch:2003} while so far (accurate)
spectroscopic redshifts are available for only about 270 galaxies
\citep{Noll:2002,Noll:2003}. Some properties of the QSO absorption
line distribution (such as the lower-than-average density of objects
at $2.8 < z < 3.1$ and an over-density near the redshift of
Q~0103-260) are also indicated in the distribution of the galaxy's
photometric redshifts. However, the accuracy of photometric redshifts
is much lower than that of the {\sc Uves} absorption line redshifts
and generally too low for a meaningful detailed comparison. The
available spectroscopic galaxy redshifts (m.e.\ about 0.002) is
adequate for the present purpose, but the number of available
redshifts is uncomfortably small for statistical
purposes. Fortunately, the FDF spectroscopic observing program is
biassed towards high redshift objects. Therefore, for the redshift
range $ 2.0 < z < 3.4 $, corresponding to (most of) the metal line
redshifts in this investigation, the present spectroscopic catalog is
$\approx $ 50~\% complete for a limiting I-Magnitude of 24.0. At
least for this $z$ range the present spectroscopic redshift catalog
can be regarded as representative sample which can be used for the
present statistical purposes.

\begin{figure}
\resizebox{\hsize}{!}{\includegraphics[angle=270]{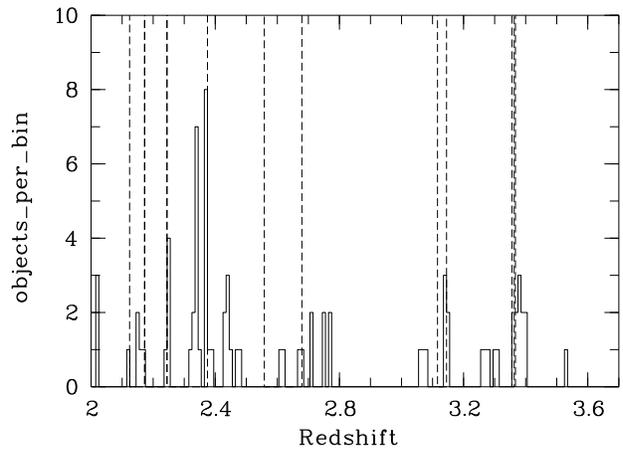}}
\caption[Histogram of FDF galaxies with $2.0 < z < 3.7$]{The redshift
distribution of FDF galaxies with accurate spectroscopic redshifts in
the range $2.0 < z < 3.7$. The broken vertical lines indicate the
redshift positions of the metal absorption systems falling into this
range. Note that the close groups of metal systems near the redshifts
2.172, 2.244, and 2.558 (cf. Table~\ref{mlinelist2}) are not
resolved on this scale and appear as single systems
in this figure.}\label{histo_FDF}
\end{figure}

Fig.~\ref{histo_FDF} presents a histogram of the $z$ distribution of
the presently available accurate spectroscopic redshifts $2.0 < z <
3.7$ in the FDF\@. Also indicated in this figure are the redshifts of
the metal absorption systems of Q~0103-260 (broken vertical lines) in
this range. In order to avoid meaningless small numbers and in view of
the expected peculiar velocities of the galaxies, we used histogram
bins of $\Delta z = 0.01$. If the redshifts of the 77 galaxies
included in Fig.~\ref{histo_FDF} were distributed randomly we would
have to expect on average 0.45 objects per bin and a Poisson
distribution of the actual number of objects per bin. While the
population of most bins is roughly consistent with this assumption,
various bins show highly significant (corresponding to 3 to 6 $\sigma
$) deviations. In particular, the clustering of galaxy redshifts at
the $z$ values 2.245, 2.34, 2.37, 2.48, 3.14 and 3.38 cannot be
explained by statistical fluctuations of a random distribution. Also
statistically significant is the total absence of objects with
redshifts $2.78 < z < 3.03$. Such strong deviations from a random
distribution of the galaxy redshifts, which have been found and
discussed before in the context of other deep fields \citep[see
e.g.][]{1996ApJ...471L...5C, 1998ApJ...492..428S,2002ARA&A..40..579G}
are obviously the result of the cosmic large-scale structure. In most
cases the galaxies included in a single redshift bin are distributed
over the whole FOV of the FDF\@. In general, the velocity
dispersion in the redshift clumps relative to the (minimum) observed
size of these structures are too large to form bound systems,
although in some bins (such as in the strong peak near $z$ = 2.37)
many objects have the same redshift within the error limits, making it
likely that these accumulations are or are becoming bound and are
precursors of galaxy clusters.

As shown by Fig.~\ref{histo_FDF}, 13 of the 16 metal system redshifts
coincide with populated galaxy redshift bins. In the absence of a
correlation (i.e.\ assuming random distributions) less than 4 such
coincidences are expected to occur. Hence, Fig.~\ref{histo_FDF}
provides not only strong evidence for the presence of large-scale
structure in the galaxy redshift distribution, but also for a strong
correlation of the galaxy distribution and the metal absorbers which
obviously trace the same large-scale structure.  The fact that no
galaxies have been found at the redshifts of the metal absorptions
system with redshift 3.116 and the two adjacent systems near $z$ =
2.558 can readily be explained by the incompleteness of our
spectroscopic survey.  More interesting is the fact that for the
conspicuous clustering of galaxy redshifts near $z$ = 2.34, in spite
of a thorough search at the expected wavelengths, no trace of
corresponding metal absorption systems could be detected. Whether
Ly$\alpha$ absorption is present at this redshift, cannot be tested
with our data since the corresponding wavelength is outside our {\sc
Uves} spectrum. Interestingly, the galaxies belonging to this
clustering differ from the other redshift groups by a pronounced
clustering tendency on the sky, with the center of their distribution
about 2 arcmin off the LOS to the QSO.

For the small spectral range, where simultaneous information on the
Ly$\alpha$ lines, the metal lines, and the galaxy redshift
distribution is available, we have included the data for all three
components in Fig.~\ref{FDF_zdistribution_2.5_3.5}. For this purpose,
the {\sc Uves} spectrum has been smoothed to a redshift resolution of 0.005
to be comparable with the resolution of the galaxy
redshifts. Moreover, the {\sc Uves} spectrum has been normalized to the
likely continuum level in the unsmoothed spectrum. The histogram of
the galaxy redshift distribution has been scaled by a factor 0.1. The
figure clearly shows the lower-than-average Ly$\alpha$ absorption and
the absence of metal line systems and galaxy redshifts in the
$z$ interval $2.80 < z < 3.05$ (corresponding to a comoving distance
of about 300 Mpc). It is also interesting to note that the strong Ly$\alpha$
absorption near $z$ = 2.76 (which does not show corresponding metal
absorption), while not exactly coinciding with the redshift of a
galaxy spectrum, nevertheless falls into a redshift region where
galaxies are observed.

We also checked our data for evidence of the weakening of Ly$\alpha$
absorption at the redshift of starburst galaxies close to the LOS
reported by \citet{Adelberger:2003}. Although some of our 
foreground galaxies are close enough to potentially produce such an effect,
our numbers are too small to provide reliable conclusions on this
question. 

\begin{figure}
\resizebox{\hsize}{!}{\includegraphics[angle=270]{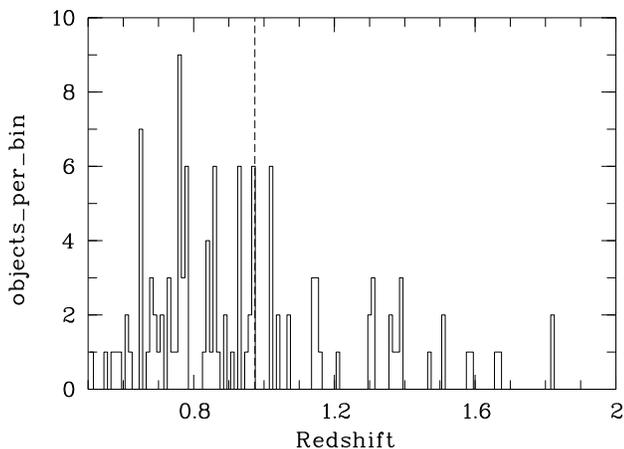}}
\caption[Positions of FDF galaxies with 0.9 $< z <$ 1.2]{Distribution
of spectroscopic galaxy redshifts in the FDF for $0.5 < z < 2.0$. The
relatively low frequency of objects between 1.0 and 2.0 is due to an
observational selection effect. The vertical broken line indicates the
position of the $z = 0.974$ Fe{\sc ii}/Mg{\sc i}/Mg{\sc ii} absorption
system described in the text.}\label{plot_FDF_09_12}
\end{figure}

As noted above, only two metal systems with (almost identical)
redshifts $<2.0$ could be reliably identified in our {\sc Uves}
spectrum.  Fig.~\ref{plot_FDF_09_12} shows the
distribution of the presently available spectroscopic redshifts and
the position of the two (on this scale coinciding) metal system
redshifts. When interpreting Fig.~\ref{plot_FDF_09_12}, one has to keep
in mind that (in contrast to Fig.~\ref{histo_FDF}) the overall
redshift distribution in Fig.~\ref{plot_FDF_09_12} is strongly
affected by an observational (negative) bias affecting galaxies in the
redshift range 1.2 -- 2.0 (cf. Noll 2002). On the other hand, the
small scale structure of the redshift distribution 
in Fig.~\ref{plot_FDF_09_12}
should be representative.  As illustrated in Fig.~\ref{plot_FDF_09_12},
the redshifts of the two detected low-$z$ systems again coincide with
a clustering of galaxy redshifts. However, the plot also shows
that there are many other such
clusterings at low redshifts for which no absorption lines were
detected.  A reason that just the systems at $z = 0.974$ are
observed could be the fact that some galaxies with this
redshift are exceptionally close to the line of sight to
Q~0103-260. One galaxy falling into the corresponding redshift bin
\citep[No. 4442 in the catalog of][]{Heidt:2003} has an angular distance
of only about 11$\arcsec$. While it still appears unlikely that gas of
this galaxy is responsible for the observed metal absorption, there are
several galaxies with similar photometric redshifts (but no
spectroscopic data) even closer to the
LOS to the QSO, which may be associated with FDF 4442. These galaxies appear
to be relatively normal late-type galaxies which could well produce
the observed redshifted interstellar absorption features.
      
\section{Conclusions}

Our analysis of the absorption lines of the quasar Q~0103-260 and a
comparison with spectroscopic redshifts of FDF galaxies 
confirm that the absorbers' redshift distribution traces well a common
large scale structure. The fact that the number density of the
Ly$\alpha$ lines shows no significant deviation from a random
distribution in the observed redshift range, while the total H{\sc i}
absorption shows variations, seems to indicate that the Ly forest
clouds in the region investigated show systematic local variations
either of their density or of their ionization level, correlated with
the galaxy density. The presence of Ly$\alpha$ clouds of high neutral
H{\sc i} column densities in the FDF region is found to be highly correlated
with strong metal line absorption and the presence of starburst
galaxies at the same redshift. This obviously suggests that hydrogen
clouds with high neutral column densities are (or have been) places of
early vigorous star formation and the enrichment of the corresponding
clouds with heavy elements. In agreement with other authors, we find
large variations in the metal line strengths and  ionization
parameters of individual Ly$\alpha$ absorbers. At the same total
Ly$\alpha$ absorption strength some clouds have strong metal lines
while others show no trace of any metal absorption.  Further studies
of larger samples will be needed to determine the physical cause of
these differences.

\begin{acknowledgements}
We would like to thank Christian Tapken and Jochen Heidt for valuable
comments on the manuscript. This research was supported by the 
German Science Foundation (DFG)
(Sonderforschungsbereich 439). 
\end{acknowledgements}

\bibliographystyle{aa}
\bibliography{frank}

\end{document}